\documentclass[%
reprint,
superscriptaddress,
amsmath,amssymb,amsthm,
aps,
soul
]{revtex4-2}

\usepackage{hyperref}
\hypersetup{
  colorlinks   = true, %Colours links instead of ugly boxes
  urlcolor     = blue, %Colour for external hyperlinks
  linkcolor    = blue, %Colour of internal links
  citecolor   = blue %Colour of citations
}
\usepackage{algorithm2e}
\SetKwComment{Comment}{/* }{ */}
\RestyleAlgo{ruled}

\usepackage{graphicx}% Include figure files
\usepackage{dcolumn}% Align table columns on decimal point
\usepackage{bm,soul}% bold math
\usepackage{xcolor}
\usepackage{braket}

\usepackage{color}

\begin{document}

\preprint{APS/123-QED}

\title{Training Classical Neural Networks by Quantum Machine Learning}% Force line breaks with \\

\author{Chen-Yu Liu}
\email{chen-yu.liu@foxconn.com}

\affiliation{Hon Hai (Foxconn) Research Institute, Taipei, Taiwan}
\affiliation{Graduate Institute of Applied Physics, National Taiwan University, Taipei, Taiwan}

\author{En-Jui Kuo}
\email{kuoenjui@umd.edu}
\affiliation{Hon Hai (Foxconn) Research Institute, Taipei, Taiwan}
\affiliation{Physics Division, National Center for Theoretical Sciences, Taipei, Taiwan}

\author{Chu-Hsuan Abraham Lin}
\affiliation{Hon Hai (Foxconn) Research Institute, Taipei, Taiwan}
\affiliation{Department of Electrical and Electronic Engineering, Imperial College London, London, UK}

\author{Sean Chen}
\affiliation{Hon Hai (Foxconn) Research Institute, Taipei, Taiwan}
\affiliation{Department of Physics, University of California San Diego, California, US}
\affiliation{Department of Mathematics, University of California San Diego, California, US}

\author{Jason Gemsun Young}
\affiliation{Industrial Technology Research Institute,  
Taipei, Taiwan}

\author{Yeong-Jar Chang}
\affiliation{Industrial Technology Research Institute,  
Taipei, Taiwan}

\author{Min-Hsiu Hsieh}
\email{min-hsiu.hsieh@foxconn.com}
\affiliation{Hon Hai (Foxconn) Research Institute, Taipei, Taiwan}

\date{\today}% It is always \today, today,
             
\begin{abstract}
In recent years, advanced deep neural networks have required a large number of parameters for training. Therefore, finding a method to reduce the number of parameters has become crucial for achieving efficient training. This work proposes a training scheme for classical neural networks (NNs) that utilizes the exponentially large Hilbert space of a quantum system. By mapping a classical NN with $M$ parameters to a quantum neural network (QNN) with $O(\text{polylog} (M))$ rotational gate angles, we can significantly reduce the number of parameters. These gate angles can be updated to train the classical NN. Unlike existing quantum machine learning (QML) methods, the results obtained from quantum computers using our approach can be directly used on classical computers. Numerical results on the MNIST and Iris datasets are presented to demonstrate the effectiveness of our approach. Additionally, we investigate the effects of deeper QNNs and the number of measurement shots for the QNN, followed by the theoretical perspective of the proposed method. This work opens a new branch of QML and offers a practical tool that can greatly enhance the influence of QML, as the trained QML results can benefit classical computing in our daily lives.
\end{abstract}

\maketitle

\section{Introduction}

% General introduction to QML
Quantum machine learning (QML) is an emergent field that utilizes the unique computational abilities of quantum systems to revolutionize the way neural networks are trained and operated. Quantum operations could be used to encode and process data into the quantum neural networks (QNNs), which could theoretically evaluate numerous possibilities simultaneously using quantum superposition and entanglement, thus accelerating the learning process \cite{qml1, qml2, qml3, qml4}. Aside from the general QNN methods, quantum kernel method utilizes quantum operations to generate kernels from the inner products of quantum states, which are then used on classical data transposed into the quantum domain \cite{qml5}. Moreover, Grover search could also be combined with QML for classification tasks, thus obtaining potential advantages \cite{gs1, gsml1}. 
% Applications of QML (QGAN, QRL, QRS, ...)
QML has exhibited diverse practical applications with significant impact, encompassing advancements in drug discovery \cite{qmlapp1}, tasks related to large-scale stellar classification \cite{qmlapp2}, natural language processing \cite{qnlp1}, recommendation systems \cite{qrs1, qrs2, qirs1, fsqc1}, and generative learning \cite{qgan1, qgan2, qgan3, qgan4, qgan5}. Despite its potential benefits, QML remains an emerging field, necessitating the resolution of several challenges before achieving widespread adoption. Key challenges include addressing issues related to the learnability \cite{qnnlearn1, qnnlearn2, qnnlearn3, qnnlearn4, qnnlearn5} and trainability \cite{qnntrain1, qnntrain2, qnntrain3, qnntrain4, qnntrain5} of QML models. QML holds significant promise as a potent tool for analyzing complex data and has the potential to bring about a revolution in various fields.
 
% Moreover, Grover's algorithm can be adapted for tasks that resemble a search problem in QML. For instance, when training a machine learning model, especially in optimization or in finding the global minimum of a loss function, Grover's search can be used to accelerate the process \red{[add ref "https://arxiv.org/pdf/2105.11603.pdf"]}. QML can improve our lives in a myriad of ways, including advancing drug discovery by rapidly analyzing molecular structures ["10.1021/acs.chemrev.8b00803 ?"], bolstering cybersecurity through enhanced encryption-breaking techniques ["10.1109/MSEC.2018.2875362" ?], and even revolutionizing climate modelling through its ability to handle extensive, complex datasets ["10.1038/srep10824" ?].

% \topic{2. Problems with QML: For the pure QML (no classical NN involving), input data encoding (gate angle encoding) makes the circuit deeper, and large data made need other preprocessing that may lose some information. For hybrid quantum classical ML, quantum computers are still required to use the trained model. Not every one have the quanutm computer, or even the access of the quantum computer through the cloud.}
% Although it seems like QML is potentially very beneficial and can impact the current status of computing, there are some inescapable issues with the current QML framework. 

In addition to issues related to learnability and trainability, the practicality of QML models represents a significant concern. In instances of pure QML devoid of classical neural networks (NNs), gate angle encoding is a common method for processing input data. However, it becomes evident that as the input data scales, the width and depth of the quantum circuit proportionally increase, hindering its accuracy in the noisy intermediate-scale quantum (NISQ) era. While classical preprocessing could mitigate this by reducing data dimension, concerns arise regarding potential information loss in such cases. To be considered practical, the dimension of data taken into the QML should be comparable to that of classical NNs. Another challenge posed against QML is that for both pure and hybrid quantum-classical machine learning (ML) models, the practical usage of the trained model at the inference stage mandates access to a quantum computer. However, quantum computers, or even cloud access to quantum computing resources, are extremely limited, introducing a significant requirement for the effective utilization of trained QML models.

% \topic{3. Similar work, to use quantum walk to train the classical neural network. This method is constrained by the nature of deep circuit for quantum walk, so that it is hard to scale up.}

To alleviate the practicality concern, a viable approach involves the utilization of quantum algorithms for training classical NNs. This strategy enables the utilization of the trained classical NN model without encountering the challenges associated with data encoding and relying on quantum computers. Prior research aligning with this perspective employs quantum walk as a searching process for classical NN parameters \cite{tcnnqc1}. Nevertheless, it is noteworthy that the Grover-like method in this context demonstrates an exponential scaling of the circuit depth.

% \topic{4. In our work, we want to provide a new perspective as well as a new QML framework that is much more practical and can address some the issue of input data encoding and inference, hopefully brings other advantages (parameter reduction). In Sec. XX, we XXX, ...}

This work aims to present a novel perspective and introduce an innovative QML framework. The proposed framework involves mapping the classical NN weights to the Hilbert space of a quantum state of QNN. This approach facilitates the tuning of the parameterized quantum state (represented by QNN), thereby enabling the adjustment of the classical NN weights. Significantly, this is achieved with a reduced parameter count of $O(\text{polylog}(M))$ compared to the classical NN with $M$ parameters. This novel approach addresses the data encoding issue of the QNN by training a classical NN with classical input and output. Moreover, once the model is trained by the QNN, inference only requires classical computers, significantly lowering the requirements for using QML results and enhancing the practicality of quantum computers in everyday life. In Sec.~\ref{sec:cnnqs}, we elucidate the mapping between quantum state measurement outcomes and classical NN weights. Subsequently, in Sec.~\ref{sec:nrd}, we delve into the numerical results and engage in a comprehensive discussion. Finally, in Sec.~\ref{sec:cfw} we sum up our results and outline the prospective avenues for future research.

\subsection{Main Results} 
% At least half page
Our main results of training classical NN weights by QML can be succinctly summarized as follows:
\begin{itemize}
    \item Shallow circuit depth of QNN: In contrast to conventional QML approaches which often incorporate data loading layers imposing constraints on input data size or requiring data compression, our methodology leverages the classical data input and output process. This provides a natural advantage, enabling a shallower quantum circuit for QNNs.

    \item Parameter reduction: In our approach only $O(\text{polylog} (M))$ parameters is required to train a classical NNs with $M$ parameters. This reduction is achieved by utilizing $N = \lceil \log_2 M \rceil$ qubits and a polynomial number of QNN layers.

    \item Inference without quantum computers: The trained model exhibits compatibility with classical hardware. In fact, the model inference relies only on classical computers, contributing to heightened practicality, particularly in light of the restricted availability of quantum computers in comparison to classical computers.
\end{itemize}

\section{Classical Neural Networks as a function of Quantum States}
\label{sec:cnnqs}

The proliferation of classical NNs equipped with millions or even billions of trainable parameters has raised concerns about the associated computational requirements. Classically, the computational power needed scales linearly with the parameters, posing a potential bottleneck as models become more sophisticated. Quantum computing, leveraging the exponential scaling of Hilbert space with the number of qubits, emerges as a promising avenue to address or alleviate this impending classical bottleneck. Motivated by this perspective, our aim is to discover a quantum state approximation of a classical NN. In this context, the quantum state's output encapsulates information about the weights, where the quantum state's parameters are exponentially fewer than those required to train a classical NN. To facilitate tunability, we express the quantum state through a QNN. By adjusting the QNN parameters, we concurrently tune the classical NN parameters, effectively constructing the classical NN as a function of the quantum state. With an $N$-qubit quantum state generating $2^N$ possible measurement outcomes, we map the probability values of diverse outcomes to classical NN weights through a scheme described later.

In this section, we initiate our discussion by elucidating the mapping from QNN measurement outcomes to classical NN weights, with 3 settings. Then the training scheme of updating the QNN corresponding to the classical NN will be presented. 

\subsection{Equivalent setting}
Consider a classical NN with $M$ weights
\begin{equation}
    \vec{\theta} = (\theta_1, \theta_2, ... , \theta_{M}),
\end{equation} we specify a quantum state $|\psi \rangle$ with $N = \lceil \log_2 M \rceil$ qubits, this guarantees the Hilbert space size $2^N \ge M$. The measurement probabilities of such quantum state in computational basis are $| \langle i | \psi \rangle |^2 \in [0,1]$, where $i \in \{1,2,...,2^N \}$. To relate $2^N$ measurement probability results to $M$ classical NN weights, the following setting is applied: 
\begin{enumerate}
    \item $2^N - M$ weights are picked randomly, with each of them be related to the average measurement probability of 2 different bases, such that
    \begin{equation}
        \theta_j = \frac{1}{2}(|\langle i | \psi \rangle|^2 + |\langle k | \psi \rangle|^2), i \neq k.
    \end{equation}
    \item After step 1, each of the remaining $2M - 2^N$ weights is related to the measurement probability of a single basis
    \begin{equation}
        \theta_j = |\langle i | \psi \rangle|^2.
    \end{equation}
\end{enumerate}
Note that even some classical NN weights are related to 2 different bases, each basis is only related to 1 weight. With the above setting, one can observed that exactly $M$ weights are mapped from measurement probabilities of $2^N$ bases.
\subsection{Positive and negative setting}
The measurement probability of $|\psi \rangle$ in the computational basis is always positive. However, a weight of the classical NN could be positive or negative. Thus, the following setting is applied to bring the negative numbers from the measurement probabilities:
\begin{eqnarray}
    \theta_j \rightarrow \theta_j \text{ for even } j, \\
    \theta_j \rightarrow -\theta_j \text{ for odd } j.
\end{eqnarray}
Such that almost half of the weights (depending on the parity of $M$) become negative and others remain positive. 

\subsection{Scaling setting}
After the above settings, the range of the classical NN weight is now become $\theta_j \in [-1, 1]$. However, as the scale of the system in consideration increases (both $M$ as well as $N$), considering the normalization condition of the quantum state
\begin{equation}
    \sum_{i=1}^{2^N} |\langle i | \psi \rangle|^2 = 1,
\end{equation}
the average probability of basis $| i \rangle$ being measured, $1/2^N$, decreases exponentially as $N$ increases. Therefore, the average magnitude of $\theta_j$ is also decreasing exponentially since $|\theta_j| \propto |\langle i | \psi \rangle |^2$. As a result, the average magnitude should be preserved under different system sizes, when designing a mapping between the quantum states and the classical NN weights. Thus, the scaling setting is designed as the following:
\begin{equation}
    \theta_j \rightarrow \gamma \tanh (\gamma 2^{N-1} \theta_j).
\end{equation}
With scaling factor $\gamma \ge 0$, the $2^N$ dependency of the scaling setting preserves the resulting average magnitude of the classical NN weights under different system sizes. The usage of the function $\tanh$ is inspired by the $\tanh$ activation function in machine learning, where the output range is $(-1, 1)$. 
% Then, we randomly choose $2^n - M$ weights mapped to 

% \subsection{Generate Classical Neural Networks weights by Quantum Neural Networks}
% \subsection{}
\subsection{Parameterized Quantum States as Quantum Neural Networks}
% \topic{Introduce the parametrization $| \psi \rangle \rightarrow | \psi (\vec{\phi}) \rangle$, and describe the EfficientSU2 ansatz (perhaps a drawing), how these gate angles (and how many? polylog M) could be changed to produce different quantum states, using classical algorithms (optimizer) COBYLA and Nelder-Mead. And most importantly, how these affect the classical NN? The data input and the cost function for training? How these data inputting different from other QML methods (gate angle encoding)? Summarize the above descriptions with the Algorithm in Algo.~\ref{alg:train}.}

In previous sections, we specified the qubit size of the quantum state $|\psi \rangle$, and the corresponding relation of the measurement probabilities to the classical NN weights. In this section, we further describe the detail of constructing $| \psi \rangle$, starting from its parametrization by introducing the rotational gate angle dependency $| \psi (\vec{\phi}) \rangle$. This parameterized quantum state could be used as the so-called QNN, with a specific ansatz that constructs the quantum circuit. The EfficientSU2 circuit ansatz is used in this work \cite{efficientsu2}, where we can specify the number of parameters that is polynomial in the number of qubits. Associate with the previous qubit number setting, the required number of parameters is $O(\text{polylog}(M))$. In Fig.~\ref{fig:scheme}(a), an example of $N = 13$ qubits EfficientSU2 with 13 layers is shown, where the circuit is initialized with Hadamard gates and consists of layers of single qubit operations spanned by SU(2) and CNOT entanglements, the random rotational angles in the example are used for illustration purpose. In the end of the circuit, all of the qubits are measured, generating $2^N$ possible outcomes. 

Following the settings in previous sections with QNN $|\psi(\vec{\phi}) \rangle$, the resulting mappings from measurement results of the QNN to the classical NN weights $\vec{\theta}$ are
\begin{eqnarray}
\label{eq:mapping1}
    \theta_{j'} = \gamma \tanh \left[\pm 2^{N-2} \gamma \left( |\langle i| \psi(\vec{\phi}) \rangle|^2+ |\langle k| \psi(\vec{\phi}) \rangle|^2 \right) \right],
\end{eqnarray}
where $j' \in \{1,2,...,2^N - M \}$, for the randomly picked indices to be related to 2 different bases. The re-indexing $j \rightarrow j'$ is for convenience to represent the amount of indices $j$, since these indices are picked randomly. The symbol $\pm$ is related to the parity of the original index $j$. And 
\begin{eqnarray}
\label{eq:mapping2}
    \theta_{j'} = \gamma \tanh \left(\pm 2^{N-1} \gamma |\langle i| \psi(\vec{\phi}) \rangle|^2 \right),
\end{eqnarray}
where $j' \in \{2^N - M +1, ..., M \}$, for the remaining weights that are related to the single basis.
\subsection{Training Classical NN by tuning QNN}
\begin{figure*}[ht]
\centering
\includegraphics[scale=0.25]{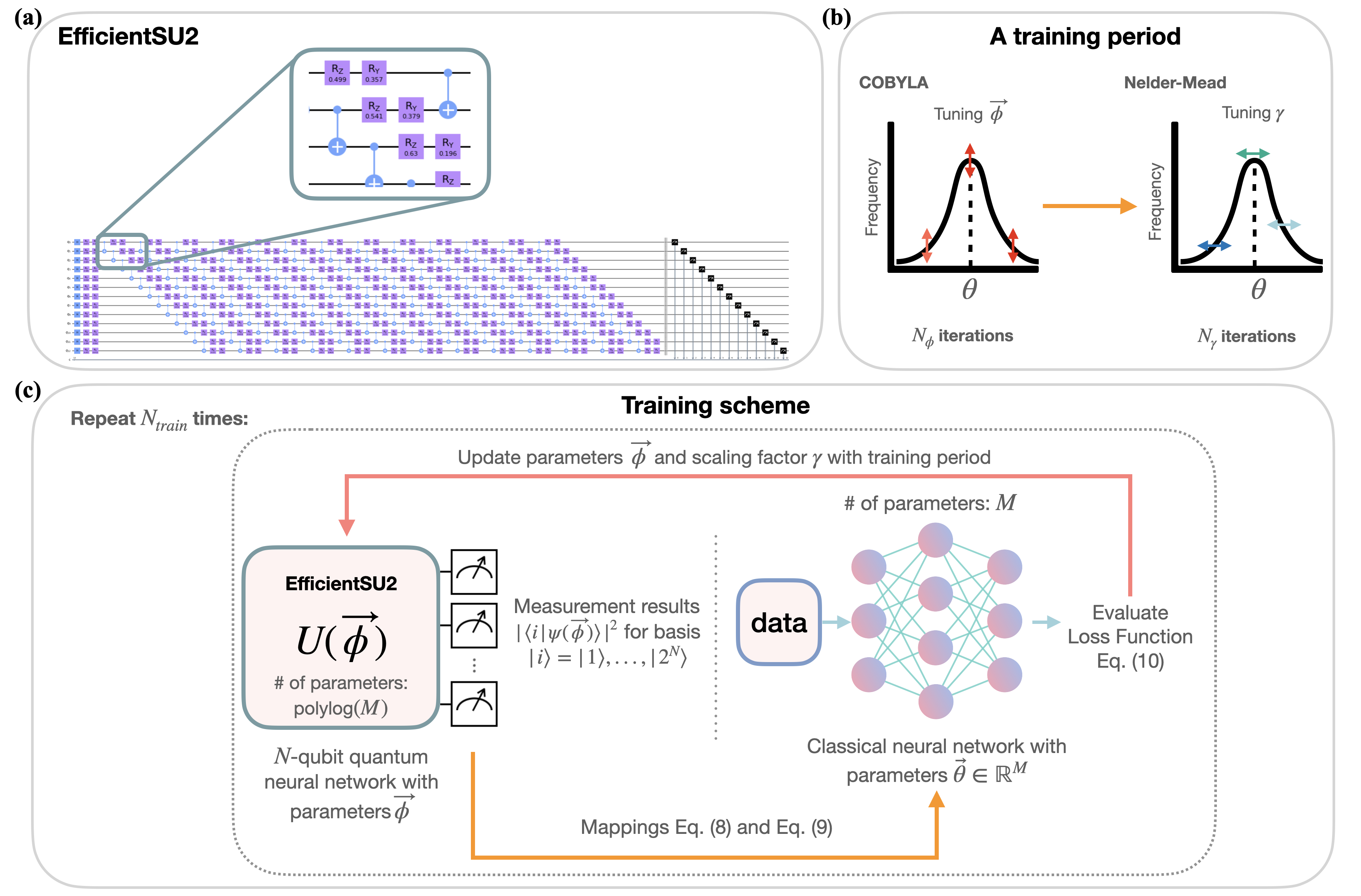}
\caption{Elements and the training scheme of the proposed approach. (a) Circuit ansatz EfficientSU2, used as QNN in this work. (b) A training period that consists of two stages, tuning the QNN parameters $\vec{\phi}$ by COBYLA algorithm, and tuning the scaling factor $\gamma$ by Nelder-Mead algorithm. Corresponding schematic histogram of the classical NN weights $\vec{\theta}$ is shown, where horizontal axis is the value of $\theta$, and vertical axis is the appearing frequency of corresponding value in $\vec{\theta}$.}
\label{fig:scheme}
\end{figure*}

For now, given a classical NN with weights $\vec{\theta} \in \mathbb{R}^M$ for some task $\mathcal{T}$, with the mappings in Eq.~(\ref{eq:mapping1}) and Eq.~(\ref{eq:mapping2}), it is possible to tune the parameters $\vec{\phi}$ of the corresponding QNN to improve the performance of classical NN on the task $\mathcal{T}$. To achieve this, the loss function and the optimization algorithm (optimizer) to tune the QNN parameters are required to be specified. 
In this work, we consider the classical NN for the classification tasks, where the data are inputted and outputted classically. The loss function to tune the QNN in order to train the corresponding classical NN is designed as
\begin{eqnarray}
    &&\text{Loss} = \text{CE}  + \frac{N_{\text{fail}}}{N_{\text{d}}}, \\
    && \text{CE} = -\frac{1}{N_{\text{d}}}\sum_{n=1}^{N_{\text{d}}}\left[y_n \log \hat{y}_n + (1-y_n)\log(1-\hat{y}_n) \right],
\end{eqnarray}
where $\text{CE}$ is the well-known Cross-entropy \cite{celoss} with $y_n$ as the true label and $\hat{y}_n$ as the predicted label by the classical NN model for the sample $n$. $N_{\text{fail}}$ is the number of failed predictions by the model, and $N_{\text{d}}$ is the total number of training data. Thus, the second term represents the rate of failed predictions. A training session in our scheme consists of $N_{\text{train}}$ training periods. In each training period, $N_{\phi}$ iterations are used to tune the parameters $\vec{\phi}$ using the COBYLA algorithm. Following that, $N_{\gamma}$ iterations are employed to adjust the scaling factor $\gamma$ using the Nelder-Mead algorithm, as illustrated in Fig.~\ref{fig:scheme}(b). The training scheme described above is summarized in Fig.~\ref{fig:scheme}(c). In contrast to ordinary QML, where QNNs typically involve data loading layers that constrain input data size or require data compression, potentially leading to information loss, our approach, as shown in Fig.~\ref{fig:scheme}(c), leverages the classical data input process, providing a natural advantage of a shallower quantum circuit for QNN. 

\section{Numerical Results and Discussion}
\label{sec:nrd}

\begin{figure*}[ht]
\centering
\includegraphics[scale=0.24]{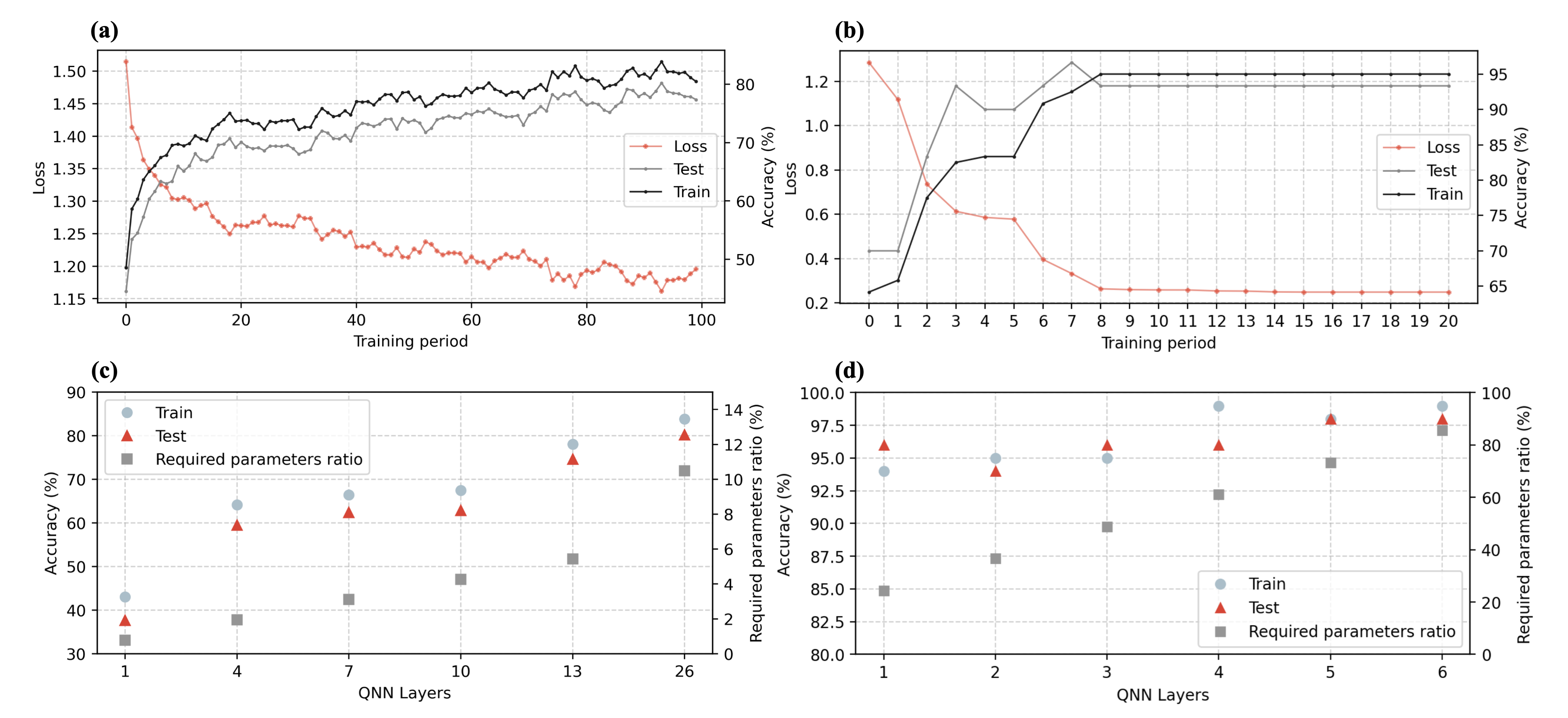}
\caption{(a) Training results for QNN layers $=26$ with the MNIST dataset. (b) Training results for QNN layers $=1$ with the Iris dataset. (c) Training results for different QNN layers with the MNIST dataset, along with the required parameters ratio for each investigated QNN layer. (d) Similar to (c) but with the Iris dataset.}
\label{fig:main_result}
\end{figure*}

\begin{figure*}[ht]
\centering
\includegraphics[scale=0.25]{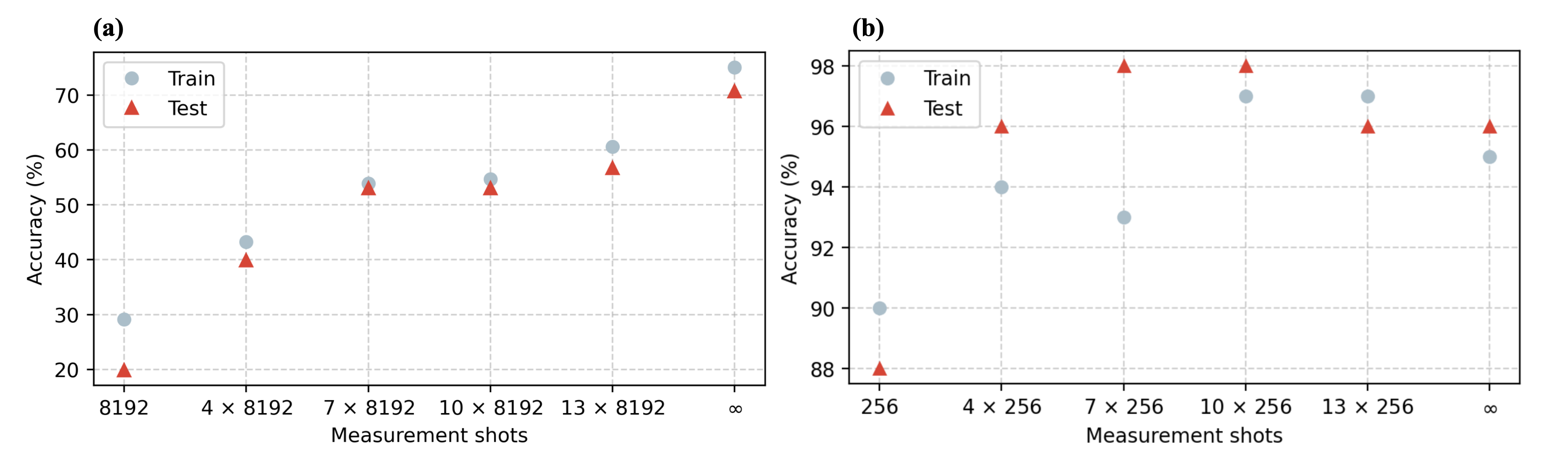}
\caption{Effects of different measurement shots. (a) MNIST dataset, where the investigated measurement shots are multiples of the corresponding Hilbert space size $2^{13} = 8192$. (b) Iris dataset, where the investigated measurement shots are multiples of the corresponding Hilbert space size $2^{8} = 256$. }
\label{fig:main_measurement}
\end{figure*}

In this section, after describing the setup of numerical experiments, we will present the training results on two different datasets. Further investigations into the effects of the number of QNN layers and that of measurement shots are also shown. 
\subsection{Setup for MNIST dataset}
\label{sec:MNIST}

In our numerical experiment conducted on the IBM Quantum simulator \cite{ibmquantum}, we initially utilize the MNIST dataset \cite{mnist} and a convolutional neural network (CNN) with $M = 6690$ parameters as our classical NN. Consequently, the required number of qubits is $\lceil \log_2 6690 \rceil = 13$. The training dataset comprises $N_\text{d} = 60000$ samples, and the testing dataset consists of $N_{\text{test}} = 10000$ samples. Within each training period, $N_{\phi} = 2000$ and $N_{\gamma} = 200$. A total of $N_{\text{train}} = 100$ training periods are conducted. The initial gate angle parameters $\vec{\phi}_{\text{init}}$ are randomly set, while the initial value of $\gamma$ is established as $\gamma_{\text{init}} = 0.3$. For now, the state vector simulation method is used, which means the number of measurement shots $= \infty$. The results for other numbers of measurement shots will be discussed later. In Fig.~\ref{fig:main_result}(a), the loss, training accuracy, and test accuracy throughout the training process are depicted for QNN layers $=26$. In this scenario, the number of QNN parameters is 728, with a required parameter ratio of $10.8\%$ compared to the original $M = 6690$.

\subsection{Setup for Iris dataset}
\label{sec:Iris}
Secondly, we also conducted on the IBM Quantum simulator using the state vector method with a number of measurement shots $=\infty$ on the Iris dataset \cite{iris}, and a single hidden layer neural network with $M = 131$ parameters is utilized. In this case, the required number of qubits is $\lceil \log_2 131 \rceil = 8$. The number of training data $N_\text{d} = 100$, and the number of testing data $N_{\text{test}} = 50$. In each training period, $N_{\phi} = 100$ and $N_{\gamma} = 10$. A total of $N_{\text{train}} = 21$ training periods are conducted. The initial gate angle parameters $\vec{\phi}_{\text{init}}$ are also set randomly, while the initial value of $\gamma$ is established as $\gamma_{\text{init}} = 0.3$. In Fig.~\ref{fig:main_result}(b), the loss, training accuracy, and test accuracy throughout the training process are presented for QNN layers $=1$. In this scenario, the number of QNN parameters is 32, with a required parameter ratio of $24.4\%$ compared to the original $M = 131$.

\subsection{Effects of QNN Layers and Measurement Shots}

With an increase in the number of QNN layers, improved expressibility is expected due to the rising number of parameters. In Fig.~\ref{fig:main_result}(c), training results for the MNIST dataset with different QNN layers $L_N \in \{1, 4, 7, 10, 13, 26\}$ are presented. Other settings, aside from QNN layers, remain consistent with the description in Sec.~\ref{sec:MNIST}. Similarly, in Fig.~\ref{fig:main_result}(d), training results for the Iris dataset with different QNN layers $L_N \in \{1, 2, 3, 4, 5, 6\}$ are also presented, with other settings following those in Sec.~\ref{sec:Iris}. It can be observed that the performance is indeed improved in both these figures. In the MNIST case, we can obtain a testing accuracy of $74.6\%$ with just a $5.38\%$ required parameters ratio and a testing accuracy of $80.21\%$ with a $10.8\%$ required parameters ratio. In the Iris dataset case, since it is a relatively simple problem compared to MNIST, with only 1 layer of QNN and a $24.4\%$ required parameters ratio, the testing accuracy is already above $95\%$. The performance slightly improves when the QNN layers increase to nearly $98\%$ when QNN layers $= 6$.

 In practice, the number of measurement shots is limited when utilizing real quantum devices. To explore the impact of different measurement shot scenarios, we conducted a comprehensive investigation. The outcomes for the MNIST dataset are illustrated in Fig.~\ref{fig:main_measurement}(a), while those for the Iris dataset are depicted in Fig.~\ref{fig:main_measurement}(b). Notably, the chosen numbers of measurement shots align with multiples of the corresponding Hilbert space size, providing a convenient framework for observing accuracy trends concerning the Hilbert space size, denoted as $2^N$. Although the precise relationship between accuracy and the requisite measurement shots remains unknown, referencing theoretical results in tomography tasks can offer insights into potential scaling behaviors. For instance, in fidelity tomography—aiming to satisfy $F(\ket{\psi}, \ket{\phi})\geq 1-\epsilon$ for the output state $\ket{\phi}$ and unknown state $\ket{\psi}$ with infidelity $\epsilon$—prior studies indicate that a sufficient number of measurement shots is $N_{\text{shot}}=O(\frac{2^N}{\epsilon} \log(\frac{2^N}{\epsilon}))$ \cite{haah2016sample}. As we contemplate investigations in larger system sizes $N$, we propose that increasing measurement shots linearly with $2^N$ is a prudent strategy for enhancing both training and testing accuracy.

\subsection{Existence and Necessity of QNN Approximation for classical NN}

After presenting the numerical results, we seek to provide the theoretical rationale for the existence of QNN approximations for classical NNs. When examining a classical NN, if the collective weights of the network give rise to a wavefunction through a mapping operation, and this wavefunction belongs to the complexity class SampBQP \cite{aaronson2016complexity, lund2017quantum, bouland2019complexity}, it logically follows that a corresponding quantum state can be generated by employing quantum circuits characterized by polynomial depth. This establishes a sufficient condition for the presence of such QNN structures.

Crucially, this condition offers compelling evidence that the quantum approximation in question cannot be replicated by a classical probabilistic Turing machine based on the quantum supremacy argument. As widely acknowledged, if SampBQP were equal to SampBPP (the complexity class defined by probability distributions that can be generated by a probabilistic Turing machine in polynomial time), it would result in the Polynomial Hierarchy collapse \cite{aaronson2011computational, lund2017quantum, bouland2019complexity}. Thus, this robustly supports the contention that SampBQP $\neq$ SampBPP. Consequently, we assert that, barring a collapse of the Polynomial Hierarchy, the incorporation of a QNN is indispensable for compressing classical NNs. It's worth noting that some tensor network \cite{montangero2018introduction, evenbly2011tensor} techniques, such as the matrix product state, can be viewed as an efficient simulation of quantum states. However, these results are limited to one dimension or geometrical local Hamiltonian.

\section{Conclusion and Future Work}
\label{sec:cfw}
% \topic{The future work including use the proposed method to fine-tune the large pretrained classical ML model.}

By using the mapping from the $2^{\lceil \log_2 M \rceil}$ measurement probabilities of QNN to the $M$ weights of classical NN, the proposed approach can significantly reduce the number of parameters to be trained. 
We demonstrated the results on the MNIST dataset and Iris dataset as the proof of concept, representing the practicality of the proposed method. 

In this study, we use the EfficientSU2 QNN ansatz, which incorporates $R_y$ and $R_z$ rotational gates along with linearly distributed CNOT gates. While there's potential for other QNN ansatzes to outperform EfficientSU2, particularly by considering qubit connectivity on real quantum hardware and devising circuits accordingly, we leave this exploration for future work. The aim for the future work is to delve into different ansatz options and assess their efficiency and effectiveness, potentially uncovering more suitable alternatives through QNN architecture search algorithms. Here, We examine the effect of number of QNN layers on the accuracy of the classification tasks in both dataset. As anticipated, our findings reveal that deeper QNNs exhibit superior performance owing to the heightened expressibility inherent in deeper circuits. It is pertinent to note that the current analysis does not incorporate noise considerations. However, in the future investigations, when the expressibility is affected by circuit noise, it is foreseeable that the performance of deeper QNNs might diminish.

Furthermore, beyond employing the precise state vector simulation for the quantum circuit, we explored shot-based simulations involving a limited number of measurement shots. Our observations reveal that the attainable accuracy in classification tasks improves as the number of measurement shots increases, with a particular emphasis on multiples of the Hilbert space size $2^N$. This outcome aligns with expectations, as certain prior theoretical studies have hinted at analogous behavior in the context of required measurement shots for quantum tomography tasks.

Currently, we have adopted a specific mapping approach; however, it is worth noting that alternative mappings could potentially enhance performance or streamline the training process. Investigating such possibilities is also designated as a future research avenue, with a focus on optimizing and exploring mappings that yield superior results in translating measurement probabilities to classical NN weights. Additionally, the optimization and learning processes employed in this study rely on non-gradient-based optimizers. A promising direction for future exploration involves studying more efficient optimization algorithms for our training task, potentially incorporating gradient calculations. Lastly, the breakthrough potential of parameter reduction on a polylogarithmic scale could revolutionize the training of large classical ML models. If quantum computing techniques can render the training of such models more efficient while allowing for inference on classical computers in our daily applications, the significance of this work would be greatly enhanced.

\acknowledgments{
EJK thanks the National Center for Theoretical Sciences of Taiwan for funding (112-2124-M-002-003). JGY and YJC acknowledge the funding provided under the ITRI EOSL 2022 project``Quantum Circuit Design and Application''.
}

\nocite{*}

\bibliography{references}% Produces the bibliography via BibTeX.

\end{document}